# Visualizing the strain evolution during the indentation of colloidal glasses

Y. Rahmani, R. Koopman, D. Denisov and P. Schall

*Van der Waals-Zeeman Institute, University of Amsterdam, Science Park 904, Amsterdam, the Netherlands*

We use an analogue of nanoindentation on a colloidal glass to elucidate the incipient plastic deformation of glasses. By tracking the motion of the individual particles in three dimensions, we visualize the strain field and glass structure during the emerging deformation. At the onset of flow, we observe a power-law distribution of strain indicating strongly correlated deformation, and reflecting a critical state of the glass. At later stages, the strain acquires a Gaussian distribution, indicating that plastic events become uncorrelated. Investigation of the glass structure using both static and dynamic measures shows a weak correlation between the structure and the emerging strain distribution. These results indicate that the onset of plasticity is governed by strong power-law correlations of strain, weakly biased by the heterogeneous glass structure.

## 1. Introduction

The investigation of incipient plasticity has been the subject of intense research in the past years. By addressing the very earliest stages of deformation, one obtains fundamental insight into the intrinsic flow mechanism of materials and the origin of plastic flow. A powerful technique in this respect is nanoindentation, in which a small tip probes a small nanometer-size volume to resolve individual relaxation events at the onset of flow [1-7]. Typically, this technique measures the applied force as a function of indentation depth to produce force-displacement curves, in which discontinuities indicate single events of deformation. The interpretation of these discontinuities is, however, difficult as the direct atomic-scale observation of the deformation remains elusive. While in crystalline materials, the incipient plasticity has been imaged on a mesoscopic level using combined indentation and transmission electron microscopy [8], such direct imaging is not possible for glasses due to the disordered amorphous structure. Rather, simulations have been used to interpret the experimental force-displacement curves [2,7]. Due to computational costs, however, these are performed at high strain rates, usually orders of magnitude higher than those of the experiments.

Colloidal crystals and glasses provide good models for conventional molecular solids [9-12]. The particles, typically around a micrometer in size, can be imaged and tracked individually in three dimensions with optical microscopy, making these systems ideal models to study the dynamics and deformation of glasses directly in real space [11,12]. Using simple equipment such as a sewing needle as indenter, indentation experiments can be performed that provide direct analogues of conventional nanoindentation [13], thus allowing insight into the atomic-scale mechanism of incipient deformation.

An important open question concerns microscopic correlations at the onset of flow. Since the early observation of amorphous materials in bubble raft models [14] and by computer simulations [15-17], the flow of glasses has been suggested to occur by localized rearrangements producing macroscopic strain. According to free volume and constitutive models, such rearrangements occur preferably at structurally weak spots that are more compliant to rearrangements. However, already early work pointed out the importance of elastic coupling between individual plastic events leading to correlations [18]. Indeed, recent simulations at zero [19] and finite temperature [20] and experimental work [21] have revealed highly correlated deformation, characterized by avalanches of localized plastic rearrangements. Such strong correlations indicate a pronounced susceptibility of the material to the applied stress field. At the particle scale, however, the deformation should be largely affected by the heterogeneous structure of the glass. As shown recently by simulations [22] and atomic force microscopy experiments [23], the heterogeneity of the amorphous structure leads to strong variations of the elastic moduli; this variation should bias rearrangements to occur at locations that are structurally weak [24,25]. This is also suggested by free volume [26] and shear transformation zone theories [27] that relate plastic deformation to the structure via a structural parameter. The question is then which of these scenarios – localized or correlated deformation or the heterogeneous structure – dominates the incipient deformation and determines the strain evolution. The direct imaging of the early stages of plasticity would allow important insight into these issues, but remains prohibitively difficult in conventional amorphous materials.

In this paper, we use indentation on a colloidal glass to elucidate the incipient deformation in glasses. By following the strain distribution with great space and time resolution, we elucidate correlations at the onset of plastic deformation. We measure the local elastic modulus from thermally induced strain fluctuations before indentation; this allows us to locate structurally weak regions in the quiescent glass. We then indent the glass and follow the evolution of the microscopic strain in detail: using correlation functions from equilibrium statistical physics, we elucidate the mechanical susceptibility of the material upon the incipient deformation. We find power-law correlations at the onset of permanent deformation, reminding of the criticality in second order equilibrium transitions, and indicating the high susceptibility of the material in the early stages of deformation. As the deformation evolves, this power-law distribution becomes overshadowed by the Gaussian strain distribution in the high-pressure zone under the needle. These results confirm and complement previous results by us, in which we reported an intriguing cascade-like mechanism of the incipient deformation [28]. We finally elucidate the relation between the emerging strain distribution and the heterogeneous glass structure. A weak but distinct correlation is observed between soft regions and regions strongly deformed upon the indentation. Our results thus highlight the predominance of strain correlations that are biased by the heterogeneous structure of the glass.

The paper is organized as follows: in paragraph 2, we describe the colloidal system and experimental setup. We then elucidate the local modulus of the glass from data taken before the indentation in paragraph 3A. The indented glass is analyzed in paragraph 3B. We finally provide conclusions in paragraph 4.

## 2. Experimental system and setup

Our colloidal glass consists of silica particles with a diameter of 1.5 µm and a polydispersity of ~5% suspended in a mixture of water (30%vol) and dimethylsulfoxide (70%vol). This solvent mixture matches the refractive index of the particles, allowing imaging of the particles deep in the bulk. A small amount of fluorescein is added to the solvent to make the particles visible as dark spots on a bright background in fluorescent imaging. We prepare a dense amorphous film by rapidly densifying the particles from a dilute suspension onto a cover slip by centrifugation. To avoid boundary-induced crystallization, the cover slip surface is coated with a 5µm-thick layer of polydisperse colloidal particles. We obtain an amorphous film about 48 µm thick, with a volume fraction of $\phi \sim 0.61$, well inside the glassy state [10]. The advantage of the colloidal system is that we can use confocal microscopy to image the individual particles in three dimensions and follow their motion in time. We do this by recording three-dimensional image stacks every minute; a single image stack takes 30 sec to acquire.

We first investigate the properties of the quiescent glass. Confocal microscopy is used to image ~30,000 particles in a 66µm × 66µm × 48µm volume and follow their motion in three dimensions during a time interval of 15min. At each time step, particle positions are determined with an accuracy of 0.02µm in the horizontal, and 0.05µm in the vertical direction [11]. The local strain is determined from the relative motion of a particle with respect to its nearest neighbours. For each particle, we determine the best affine transformation tensor $\alpha$ that transforms the nearest neighbour vectors over the time interval [12,17]. The symmetric part of $\alpha$ is the strain tensor of the particle under consideration. We thus compute the thermally induced strain distribution in the quiescent glass.

We then indent the amorphous colloidal film using a sewing needle. The needle has an almost hemispherical tip with a radius of 38 µm and is attached to a piezoelectric translation stage to move it at a slow and well-controlled speed. We lower the needle at a constant speed of 2.9µm/h into the amorphous sediment, and acquire image stacks every minute for a total duration of 25min. We again determine the strain distribution from the motion of a particle relative to its nearest neighbours. We compute both the accumulated strain (with respect to the first frame) and the instantaneous strain (with respect to the previous frame). The time $t = 0$ is defined as the instance when the needle touches the amorphous film, i.e. the upper layer of particles.

## 3. Results

### 3A. Structure and dynamics of the quiescent glass

The properties of the quiescent glass are illustrated in Fig. 1. The pair correlation function depicted in Fig. 1a shows the characteristic short range order of a liquid with a strong nearest neighbor peak indicating the dense packing of a glass. The mean-square displacement of the particles (Fig. 1b) reveals the characteristic plateau of a glass indicating the dynamic arrest of the suspension on the experimental time scale. Nevertheless, strong fluctuations of the particles occur around their rest positions: the particles exhibit short-time diffusion in the cage constituted by their neighbors. This thermal motion leads to fluctuations in the particle positions that manifest as strain fluctuations. To make these visible, for each particle, we determine the motion relative to its nearest neighbors between two subsequent frames, and we compute the local strain tensor using affine fitting. We show reconstructions of the local shear strain $\varepsilon_{xy}$ in subsequent frames in Figs. 1c and d. Regions of alternating strain are clearly visible (dotted circles); these demarcate fluctuations of strain. However, no permanent strain builds up as shown by plotting the mean-square strain as a function of time in Fig. 1b: A plateau similar to that of the mean-square displacement indicates that no strain accumulates over time.

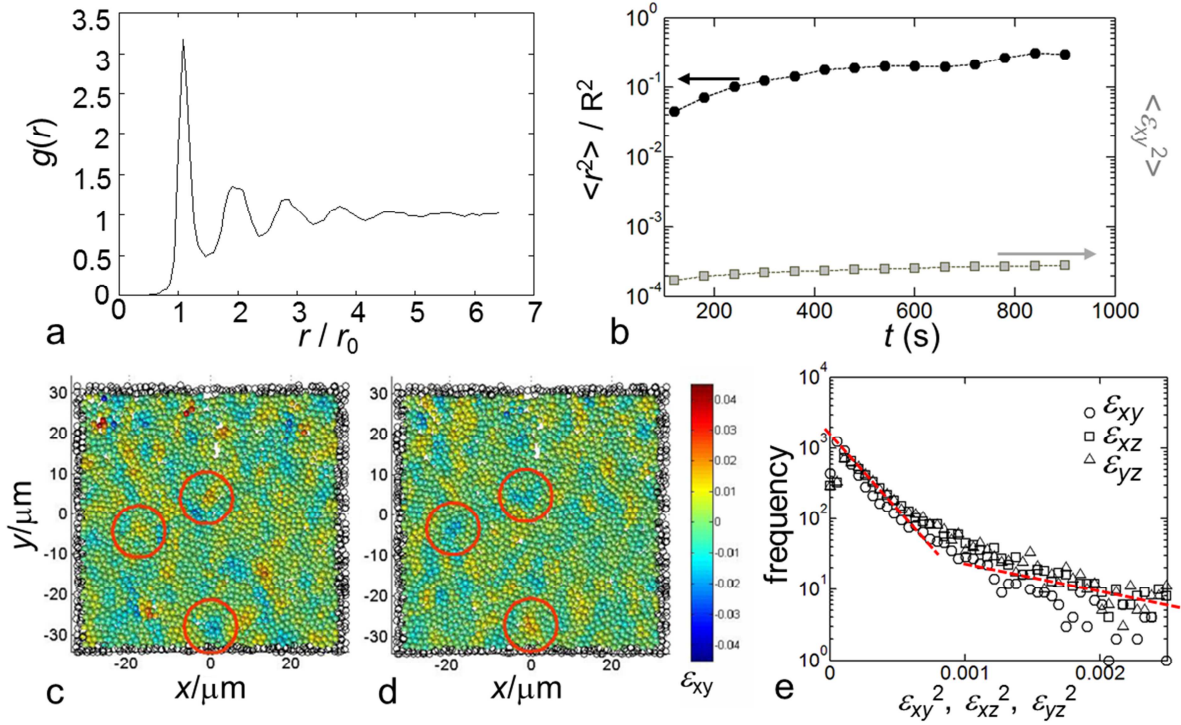

**Fig. 1 Structure and dynamics of the quiescent glass**
(a) Pair correlation function and (b) mean-square displacement and mean-square shear strain as a function of time. The pronounced nearest neighbor peak of the pair correlation function and the plateau of the mean-square displacement demonstrate glass-like properties. (c,d) 5μm thick reconstructions show subsequent strain distributions in the quiescent glass, 9μm below the surface of the amorphous film. Circles indicate zones of alternating strain. (e) Frequency of strain as a function of strain magnitude for the three shear strain components $\varepsilon_{xy}$, $\varepsilon_{xz}$, and $\varepsilon_{yz}$.

We use these thermally induced strain fluctuations to determine the elastic modulus of the amorphous film. Because of local thermal equilibrium, we expect that strain energies follow a Boltzmann distribution. We can thus determine the elastic modulus of the amorphous film from the probability distribution of strain magnitudes. Assuming the glass to be an isotropic elastic solid with shear modulus, $\mu$, the elastic energy associated with the shear component $\varepsilon_{ij}$ is $E_{ij} = (1/2)\mu\varepsilon_{ij}^2 V_\varepsilon$, where $V_\varepsilon$ is the volume over which the strain is computed. Hence, shear strains should occur with probability $P(\varepsilon_{ij}) \propto \exp(-\mu\varepsilon_{ij}^2 V_\varepsilon / 2kT)$. To test this prediction, we plot the relative frequency of strains as a function of $\varepsilon_{ij}^2$ in a half-logarithmic representation in Fig. 1e. In this representation, the slope of the data indicates the shear modulus of the amorphous film in units of $(2kT/V_\varepsilon)$. A single slope is expected for a homogeneous material. Instead of a single slope, however, we notice that the data can be fitted by a range of slopes, as indicated by the dotted lines demarcating the highest and lowest slopes. The range of slopes suggests that the modulus is not homogeneous, but instead varies across the film. The two slopes yield moduli of $\mu_{max} = 85\ kT/r^3$ and $\mu_{min} = 18\ kT/r^3$, corresponding to 0.85 and 0.18 Pa, respectively. The heterogeneous glass structure thus appears to result in a spatially varying modulus, with magnitudes between $\mu_{max}$ and $\mu_{min}$. To confirm this hypothesis, we look at the distribution of strain in more detail. We notice that the magnitude of strain fluctuations is indeed not homogeneous across the sample: Some regions exhibit stronger fluctuations than others, and they do so persistently over time. Stronger fluctuations demarcate structurally weaker regions. Assuming that in local thermal equilibrium, strain fluctuations are excited on average with thermal energy $kT$, for every shear strain component $\varepsilon_{ij}$, fluctuations will be observed on average with mean-square amplitude according to $kT = (1/2)\mu <\varepsilon_{ij}^2>$. This allows us to measure the local modulus from the average variance of the local strain. To calculate $<\varepsilon_{ij}^2>$, we use the time average of $\varepsilon_{ij}^2$ over the recorded 15 time steps; no permanent strain accumulates during this time interval (see Fig 1b). We thus obtain the local modulus for each particle and its surrounding, and we smoothen the results by averaging the modulus of the particle with that of the particle's nearest neighbors. The resulting map of the local modulus, shown in Fig. 2a, reveals strong heterogeneity; the maximum and minimum values are indeed in good agreement with the slopes in fig. 1e. This heterogeneity is also in qualitative agreement with computer simulations of Lennard-Jones glasses [22] and measurements on metallic glasses by atomic force microscopy [23].

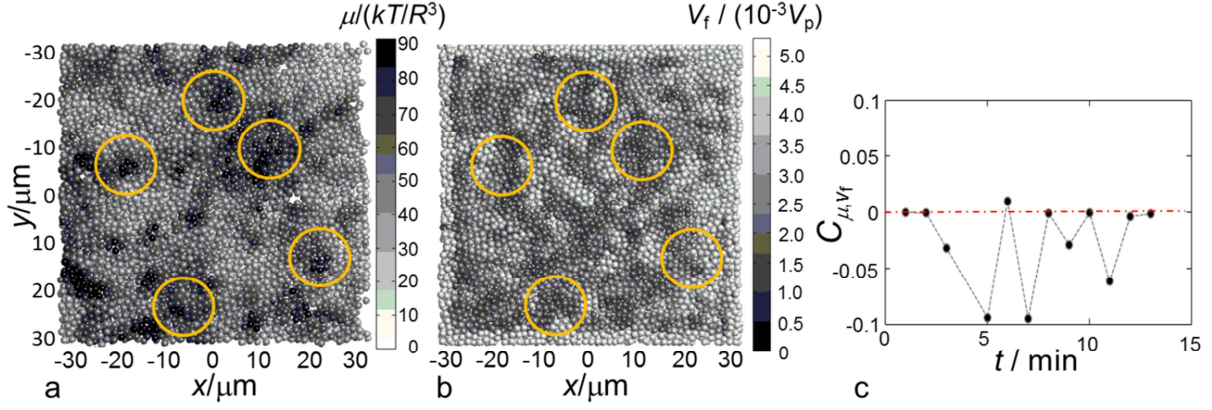

**Fig. 2 Heterogeneity of the elastic modulus and free volume**
(a) Reconstruction of the local elastic modulus in a 5µm thick section of the glass, 9µm below the surface. (b) Free volume distribution in the same section as depicted in (a). Circles indicate corresponding regions with high local modulus and low free volume. (c) Normalized correlation between the local modulus and the free volume as a function of time.

To explore the link between the microscopically varying modulus and the heterogeneous glass structure, we determine the free volume of the particles. The free volume indicates the space within which the center of a particle can move without moving its neighbors. To estimate the free volume, we construct Voronoi cells that include all points closer to the particle than to any other particle. We then move the Voronoi faces inwards by a distance equal to the particle diameter. The remaining small volume gives an estimate of the space in which the center of the particle can move [29]. We show a grayscale representation of the resulting free volume distribution in Fig. 2b. Its heterogeneity reflects the strongly varying environment of the particles. By comparing Figs. 2a and b, we notice a weak negative correlation between the local modulus and the free volume: stiffer regions (high local modulus) tend to have smaller amount of free volume, and vice versa. We quantify this relation by determining the normalized correlation coefficient,

$$C_{\mu,V_f} = \frac{\sum_i (\mu_i - \langle\mu\rangle)(V_{f,i} - \langle V_f\rangle)}{\sqrt{\sum_i (\mu_i - \langle\mu\rangle)^2 \cdot \sum_i (V_{f,i} - \langle V_f\rangle)^2}} \quad (1)$$

that correlates fluctuations of the particles´ local shear modulus, $\mu$, and free volume, $V_f$. To evaluate eq. (1), we use the time average of $\mu$, and correlate it with instantaneous distributions of $V_f$ obtained from the 15 individual snapshots. The resulting correlation coefficient as a function of time is shown in Fig. 2c. This correlation coefficient fluctuates between 0 and 0.1, showing on average a weak negative correlation, in agreement with the expected reciprocal relation between the modulus and free volume. These results suggest indeed a weak link between the heterogeneous modulus and the heterogeneous amorphous structure.

### 3B. Strain evolution in the indented glass

After elucidating the properties of the quiescent amorphous film, we probe its incipient deformation by indentation. We indent the glass using a sewing needle that we push slowly into the amorphous sediment; the indentation speed of 2.9μm/h is sufficiently slow so that thermally activated rearrangements can occur. We image individual particles in a 66μm by 66μm by 45μm volume below the tip and follow their motion in three dimensions using confocal microscopy. Again, we determine the local strain from the motion of particles with respect to their nearest neighbours. Reconstructions of the normal strain component $\varepsilon_{zz}$ are shown in Fig. 3a-c. The increasing indentation pressure is clearly visible as emerging negative (compressive) normal strain below the needle. We also determine the maximum shear strain defined by $\varepsilon_{max} = |\varepsilon_1 - \varepsilon_2|/2$, where $\varepsilon_1$ and $\varepsilon_2$ are the largest and smallest eigenvectors of the strain tensor. The maximum shear strain is an invariant of the strain tensor; it reflects the shear strain acting along the principal axes of the strain tensor and provides a good measure of the local shear deformation in amorphous materials [30]. Its evolution is shown in Fig. 3d-f. Yellow and orange particles indicate zones of high local $\varepsilon_{max}$; with progressing indentation, these accumulate below the needle as expected. The position of the maximum of $\varepsilon_{max}$ is also in agreement with continuum elasticity that predicts that the maximum occurs at a distance $h_{max}$ of about half the contact radius below the specimen surface [30-32]. To show this, we determine the contact radius $r_c = 22$μm at $t = 21$min directly from the three-dimensional images and indicate $h_{max}$ in a vertical slice through the indented glass in Fig. 3g. The height of the strain maximum is indeed in good agreement with the height predicted by continuum elasticity, while its lateral position is somewhat off center. We further notice that in addition to the central maximum, high strain persists even further away. In fact, a closer look at Figs. 3d-f shows an interesting structure of deformation with high strain zones forming at specific distances to the center (concentric dotted circles) [28]. This structure contrasts with the smooth, symmetric strain distribution of elastic deformation [30-32], and indicates the emerging plastic flow [28]. We note that this structure of deformation is different from the sharp slip events typically observed in the indentation of metallic glasses [3,4,7]: The low strain rates applied here address the regime of homogeneous deformation, while most indentation on metallic glass has addressed the inhomogeneous regime [4]. The low strain rate and long time scale of the colloidal glass studied here allow us to investigate the generic mechanism of the onset of homogeneous deformation.

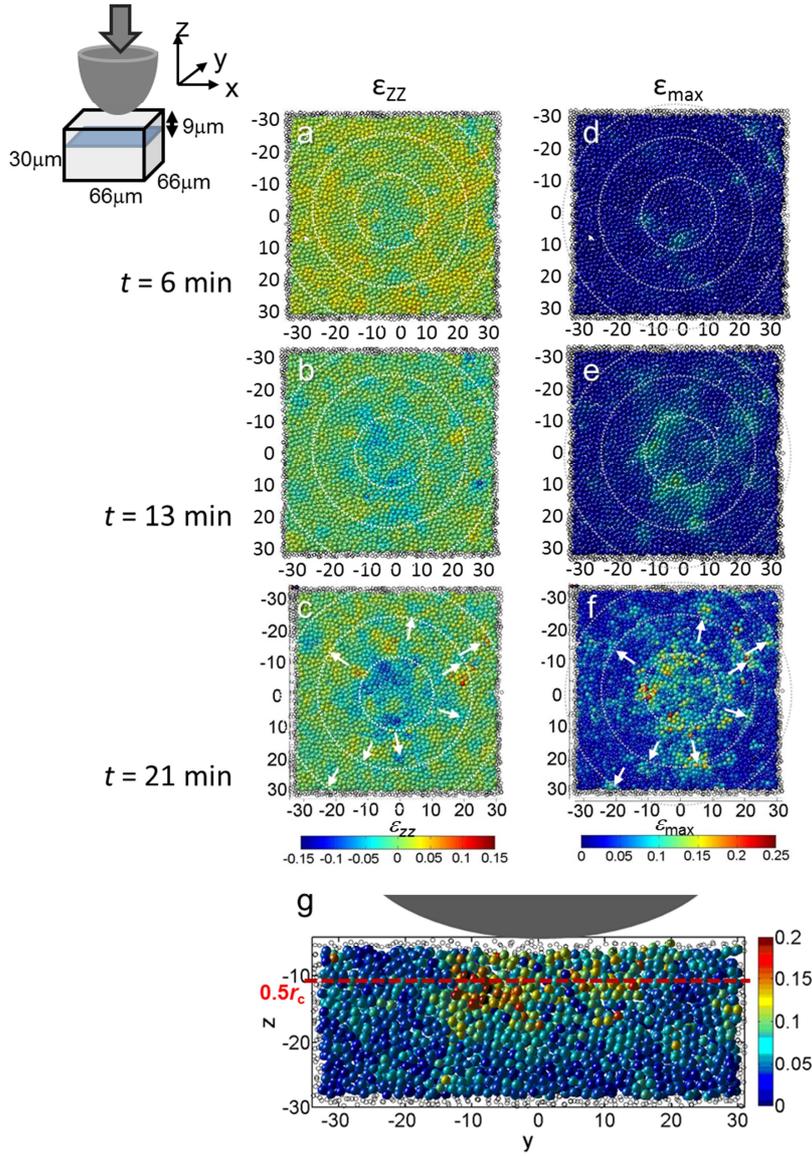

**Fig. 3: Strain evolution during indentation**
Reconstructions of the normal strain $\varepsilon_{zz}$ (a-c), and the maximum shear strain $\varepsilon_{max}$ (d-f) after 6, 13, and 21 min of indentation. The panels show 5μm thick horizontal sections, 9μm below the surface. Concentric dotted circles indicate distances of $r = 10$, 21 and 40μm to the center. (g) Reconstructions of $\varepsilon_{max}$ at $t=21$min in a 5μm thick vertical section at $x = 0$μm.

To investigate the strain distribution at the onset of plastic flow, we focus on the early stages of the indentation and use spatial correlation functions to measure the coherence and range of typical strain fluctuations. We define [21]

$$C_\varepsilon(\Delta r) = \frac{<\varepsilon(r)\varepsilon(r+\Delta r)> - <\varepsilon(r)>^2}{<\varepsilon(r)^2> - <\varepsilon(r)>^2}, \tag{2}$$

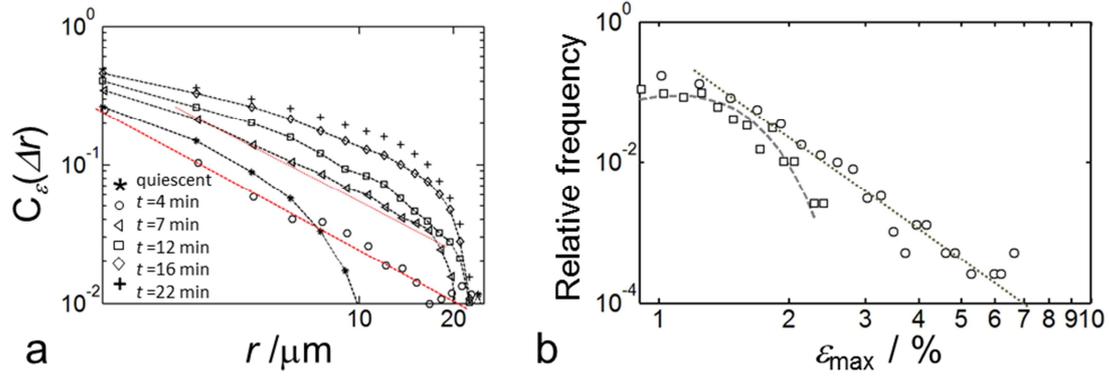

**Fig. 4 Strain correlations and distribution function**
(a) Spatial correlation of the fluctuations of the maximum shear strain for the quiescent glass (stars), and after 4 (circles), 7 (triangles), 12 (squares), 16 (diamonds) and 22 min (crosses) of indentation. Dashed lines indicate power-law correlations at early stages of indentation. (b) Relative frequency of strain values as a function of strain magnitude for particles inside (squares) and outside the central high-strain zone (circles).

which correlates strain at locations separated by $\Delta r$. Such spatial correlation functions are used in second order transitions to measure the increasing susceptibility of the material to external fields. We apply this formalism here to elucidate the mechanical susceptibility of the glass under the applied deformation. Indeed, this analysis reveals an interesting behaviour of the incipient microscopic strain as shown in Fig. 4a: At the early stages of indentation, the correlation function acquires a power-law distribution (dotted red lines), indicating long-range correlations and a high susceptibility of the system to the applied (mechanical) field. At this early stage, the applied pressure is merely strong enough to induce the first (highly correlated) rearrangements. At later stages of indentation (upper curves) the strain correlations become dominated by the strong strain in the centre, leading to higher correlations, but faster decay. Similarly, in the quiescent glass, strain correlations decay quickly, without power-law signature. This is because the motion of particles is restricted by nearest-neighbour cages leading to localized fluctuations only. On the other hand, at later stages of indentation, the strong pressure in the centre allows rearrangements to occur independently of each other. In between, where the applied pressure is just sufficient to trigger rearrangements, these occur cooperatively, resulting in long-range correlations. To further confirm this interpretation, we investigate strain distributions in and outside the central high-pressure zone. The corresponding probability distributions are shown in Fig. 4b. Two characteristically different distributions are observed: particles in the centre exhibit a Gaussian distribution, indicating deformation is localized and uncorrelated. In contrast, particles outside the centre exhibit a power-law strain distribution, indicating strongly correlated deformation. This confirms our interpretation that in the central high-pressure zone, rearrangements occur independently of each other, while under the smaller shear stress outside the centre deformation occurs in a correlated manner. At the same time, the glass structure remains essentially robust. To illustrate this, we plot the pair correlation function and free volume distribution of particles in and outside the centre in Fig. 5, where we also include distributions of the quiescent glass before indentation. No significant difference is visible in $g(r)$, while a small difference occurs in the free volume distribution, indicating a potentially interesting effect of the applied deformation.

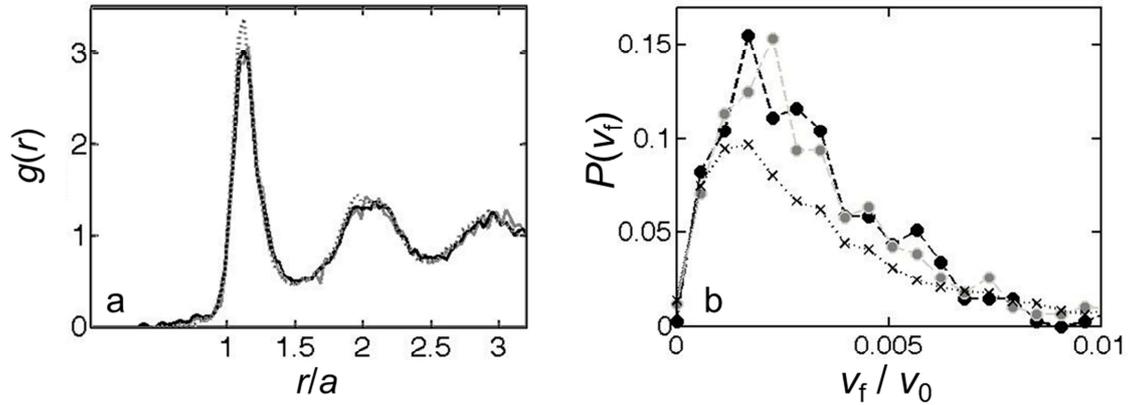

**Fig. 5 Structure of the indented glass**
Pair correlation function (a) and distribution of free volume (b) for particles inside (black line and dots) and outside the high pressure zone (gray line and dots), as well as in the quiescent glass before indentation (dotted line and crosses).

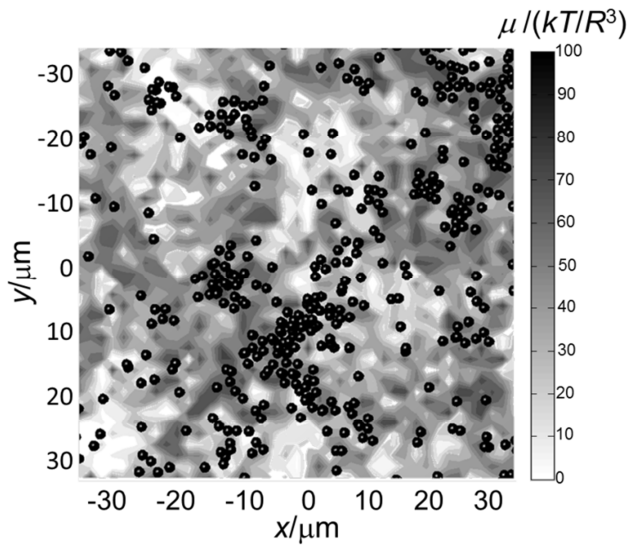

**Fig. 6 Correlation: local modulus - shear strain**
High-strain particles ($\varepsilon_{max} > 0.05$ after 6min of indentation) superimposed on a contour plot of the local shear modulus (determined before indentation). High strain particles show a preference to occur in regions of low modulus.

In this context it is interesting to investigate coupling between the structure and the emerging strain. For example, the inhomogeneous elastic modulus could bias the emerging strain to occur in structurally weak regions. We therefore investigate correlations between soft regions and those of strong deformation. To give a visual impression of this relation, we show a contour plot of the local modulus with overlayed high-strain particles in Fig. 6. Indeed, high-strain particles show some tendency to occur in or close to soft regions, in agreement with earlier work [24,25]. We determine this relationship quantitatively by calculating the correlation coefficient

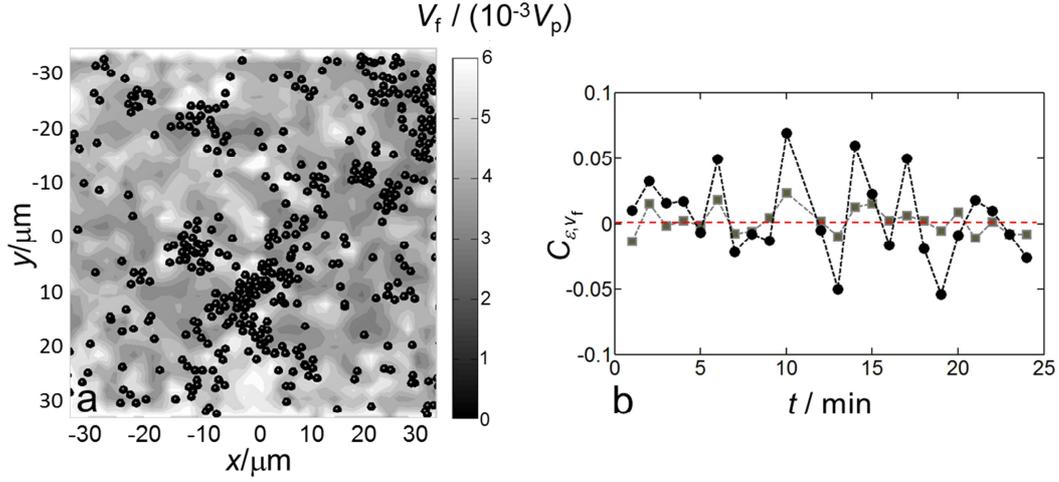

**Fig. 7 Correlation: free volume – shear strain**
(a) High-strain particles ($\varepsilon_{max} > 0.05$ after 6min of indentation) superimposed on a contour plot of the local free volume. (b) Correlation between shear strain and free volume as a function of time. Gray squares indicate all particles, and black dots indicate only high-strain particles ($\varepsilon_{max} > 0.05$). No clear correlation is observed.

$$C_{\mu,\varepsilon_{max}} = \frac{\sum_i (\mu_i - \langle \mu \rangle)(\varepsilon_{max,i} - \langle \varepsilon_{max} \rangle)}{\sqrt{\sum_i (\mu_i - \langle \mu \rangle)^2 \cdot \sum_N (\varepsilon_{max,i} - \langle \varepsilon_{max} \rangle)^2}} \quad (3)$$

that correlates fluctuations of the particles' local modulus with fluctuations of the indentation-induced strain. We evaluate eq. 3 for all high-strain particles shown in Fig. 6, and obtain $C_{\mu,\varepsilon_{max}} \sim -0.15$, indicating significant correlation. The negative sign indicates that larger strain occurs preferentially at locations exhibiting smaller local modulus and vice versa, as expected. We thus confirm that rearrangements tend to occur in weak regions. Hence, while the overall strain evolution is governed by strong strain correlations, it is biased towards structurally weak regions, i.e. regions with a low local modulus.

Again, it is interesting to investigate whether this structural bias extends directly to properties of the static structure. A number of theories [26,27] suggest a correlation between local rearrangements and structural parameters such as the free volume: a larger amount of free volume makes a region rearrange more easily and thus makes it more susceptible to deformation. This connection is already suggested by Fig. 2, but it is not clear whether this correlation persists during the initial stages of deformation. To investigate this relationship, we again provide a visual impression by overlaying high-strain particles onto a contour map of the free volume in Fig. 7. This plot suggests maybe a weak preference for high-strain particles to appear in regions of high free volume. To test this relation quantitatively, we define a correlation coefficient similar to eq. 3, however with $\mu$ replaced by $V_f$. The resulting correlation function, $C_{V_f,\varepsilon_{max}}$, is shown as a function of time in Fig. 7b. Grey squares indicate the correlation coefficient for all particles, while black circles indicate the correlation coefficient for high-strain particles only. The data does not show a clear correlation. The weak trend towards positive correlation is overshadowed by strong fluctuations.

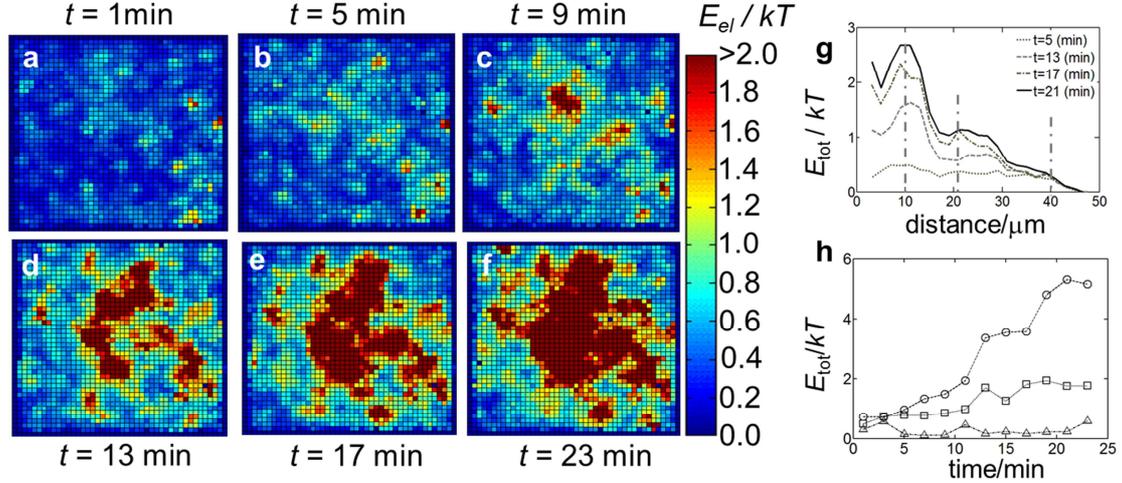

**Fig. 8: Evolution of the total elastic energy**
(a-f) Reconstructions of the total strain energy computed from all strain components of the particles. The time series reveals the strong spatial and temporal heterogeneity of the deformation. (g) Angle-averaged strain energy as a function of distance from the center after 5, 13, 17 and 21 min of indentation. Dash-dotted vertical lines delineate distances of high activity. (h) Total energy as a function of time at the distances indicated by dash-dotted lines in (g).

The hallmark of the incipient deformation is thus the strongly correlated strain. This strain correlation arises from the elastic coupling between transforming regions, mediated by their characteristic strain fields [21]. The resulting strain evolution is complex, and can lead to interesting space-time structures. We elucidate these structures in more detail by following the total elastic energy released during the indentation. Assuming that the simple linear elastic approximation holds, we can compute the elastic energy of the neighbourhood of a particle from its strain tensor; this should be a reasonable approximation for strains smaller than ~0.1, i.e. for most of the particles; the approximation is less good for strain values larger than 0.1, which correspond to the highest strain values observed here. In the linear elastic approximation, the total strain energy density is $E_{tot} = (1/2)\mu(2\varepsilon_{ij}^2 + \lambda\varepsilon_{kk}^2)$ [33], where the squared $\varepsilon_{ij}^2$ and $\varepsilon_{kk}^2$ denote the sum over all components. We replace the Lamé constant $\lambda$ by $2\nu\mu/(1-2\nu)$ with the Poisson ratio, $\nu = 1/3$. We thus compute the elastic energy for all particles from their time-dependent strain, and show its evolution in Fig. 8a-f. These reconstructions demonstrate the strong spatial and temporal heterogeneity of the incipient deformation. Already at early times, high-strain zones span the entire field of view; these zones develop into a fractal-like structure at later times, with high activity occurring at specific distances to the centre. The time evolution is demonstrated in Figs. 8g and h. In fig. 8g, we plot the angle-averaged elastic energy as a function of radial distance for increasing time intervals. An interesting structure emerges with high activity concentrating at characteristic distances to the centre; these distances correspond to the dashed circles in Fig. 3. To elucidate the time evolution in these high-activity regions, we follow the strain as a function of time in thin shells around these characteristic distances, see Fig. 8h. Distinct bursts are observed in the time evolution; sudden jumps occur after characteristic time intervals, demonstrating temporal intermittency and interesting space-time correlations.

## 4. Conclusion

We have investigated the indentation of a colloidal glass by direct real-space imaging of the microscopic strain distribution. We find that the onset of plastic deformation is dominated by strongly correlated strain caused by the glass' elasticity. At the onset of permanent deformation, strain fluctuations acquire power-law correlations, indicating a high susceptibility of the material to the applied mechanical field, and a critical state at the onset of flow. These correlations are weakly biased by the underlying heterogeneous glass structure. Correlation analysis suggests that there is a weak connection between the emerging strain and the underlying heterogeneous modulus, while direct correlations with structural measures such as the free volume are insignificant. The observed power-law correlations indicate that the transition from reversible elastic to irreversible plastic deformation has signatures of a critical point. These strain fluctuations, however, reflect a dynamic phenomenon, the description of which should include time as explicit variable. Our time-resolved analysis shows that indeed intermittency occurs not only in space, but also along the time dimension. Understanding and describing the full spatio-temporal structure of the emerging deformation is a central challenge for future theories of the incipient deformation of glasses.

**Acknowledgement**

This work is part of the research program of FOM (Stichting voor Fundamenteel Onderzoek der Materie), which is financially supported by NWO (Nederlandse Organisatie voor Wetenschappelijk Onderzoek). P.S. acknowledges financial support by a VIDI fellowship from NWO. We thank R. Zargar for supplying his code for measuring the free volume.